\documentclass[]{spie}  

\usepackage{amsmath,amsfonts,amssymb}
\usepackage{graphicx}
\usepackage[colorlinks=true, allcolors=blue]{hyperref}

\title{Investigation of InAs--based devices for topological applications}

\author[a]{Matteo Carrega}
\author[a]{Stefano Guiducci}
\author[a]{Andrea Iorio}
\author[a]{Lennart Bours}
\author[a]{Elia Strambini}
\author[b]{Giorgio Biasiol}
\author[a,c]{Mirko Rocci}
\author[a]{Valentina Zannier}
\author[a]{Lucia Sorba}
\author[a]{Fabio Beltram}
\author[a,d]{Stefano Roddaro}
\author[a]{Francesco Giazotto}
\author[a]{Stefan Heun}
\affil[a]{NEST, Istituto Nanoscienze-CNR and Scuola Normale Superiore, I-56127 Pisa, Italy}
\affil[b]{IOM CNR, Laboratorio TASC, Area Science Park, 34149 Trieste, Italy}
\affil[c]{Francis Bitter Magnet Laboratory and Plasma Science and Fusion Center, Massachusetts Institute of Technology, Cambridge, MA 02139, USA}

\affil[d]{Dipartimento di Fisica, Università di Pisa, Largo Bruno Pontecorvo 3, I-56127 Pisa, Italy}

\authorinfo{Further author information: (Send correspondence to M.C.)\\M.C.: E-mail: matteo.carrega@nano.cnr.it\\S.G.: Deceased.}

\begin{document} 
\maketitle

\begin{abstract}
Hybrid superconductor/semiconductor devices constitute a powerful platform to investigate the emergence of new topological state of matter.
Among all possible semiconductor materials, InAs represents a promising choice, owing to its high quality, large g-factor and spin--orbit component. Here, we report on InAs-based devices both in one--dimensional and two--dimensional configurations. In the former, low-temperature measurements on a suspended nanowire are presented, inspecting the intrinsic spin--orbit contribution of the system. In the latter, Josephson Junctions between two Nb contacts comprising an InAs quantum well are investigated. Supercurrent flow is reported, with Nb critical temperature up to $T_c\sim 8$~K. Multiple Andreev reflection signals are observed in the dissipative regime. In both systems, we show that the presence of external gates represents a useful knob, allowing for wide tunability and control of device properties, such as spin--orbit coherence length or supercurrent amplitude. 
\end{abstract}


\section{INTRODUCTION}
\label{sec:intro}  
  
Recently, promising platforms based on superconductor/semiconductor systems have attracted a lot of interest,  especially for the possible emergence of new topological state of matter, paving the way for new quantum technologies with intrinsic robustness\cite{kitaev_unpaired_2001, oreg_helical_2010, Mourik2012,Albrecht2016,Mong2014,Alicea2016, Wan2015,Tiira2017,Wickramasinghe2018}. Several proposals, based on the combination of superconductivity, spin--orbit coupling, and external magnetic fields, have been introduced, based on which quasiparticles like Majorana fermions\cite{kitaev_unpaired_2001, oreg_helical_2010, Alicea2016}, or more exotic parafermions\cite{Clarke2013,Lindner2012,Clarke2014, Blasi_2012, Mazza_2018, Rossini_2019}, may form. These pioneering predictions have triggered great experimental efforts, and important milestones have already been reported\cite{Mourik2012,Albrecht2016, Tiira2017}. Nonetheless, a complete understanding of hybrid structures and a smoking--gun--signature of topological quasiparticles is still lacking. It is thus important to inspect different platforms, e.g.~exploiting nanowire--based setups or hybrid junctions with two--dimensional electron gas (2DEG) systems, to seek for the best candidate and to find the optimal conditions for new and scalable architectures.

Among all semiconductor materials, InAs plays a prominent role due to its low effective mass, strong spin--orbit coupling, and high Land{\'e} g-factor\cite{Takayanagi1995,Shabani2016,Kjaergaard2016,Kjaergaard2017,Goffman2017,Sestoft2018}. Moreover, hybrid superconductor/semiconductor devices require high quality contacts with low normal/superconductor (N/S) interface resistance, to guarantee a robust proximity effect, and eventually large electron mean free path (MFP)\cite{Goffman2017,Casparis2017}. It has been shown that InAs represents the ideal choice for building hybrid devices, owing to the lack of a Schottky barrier at the interface with the metal, combined with the small effective mass and large spin--orbit coupling of this semiconductor.\cite{Takayanagi1995,Shabani2016,Kjaergaard2016,Kjaergaard2017,Goffman2017,Sestoft2018} Epitaxial Al/InAs heterostructures were realized that show an exceptionally transparent superconductor-semiconductor interface and the presence of a hard superconducting gap induced by the proximity effect in a semiconductor\cite{Goffman2017, Casparis2017}. These features, combined with the possibility of external control by means of field effect, playing with side gates properly patterned close to the devices, make this material a favourable candidate for a broad range of applications, from spintronics to superconducting qubits and topological quantum computation\cite{golovach_electric-dipole-induced_2006, nowack_coherent_2007, manchon_new_2015, desrat_anticrossings_2005, Nayak2008}. 

Here, we report on two main research lines covering the physics of InAs-based devices and hybrid structures and how electrostatic gates can be exploited both to tune device properties and induce additional fields. 
First we show the realization of suspended InAs nanowires, and their characterization by means of low-temperature transport measurements\cite{iorio_nanolett_2019}. This geometry allows for the investigation of the intrinsic spin--orbit component which, generally, can be strongly affected and modified by the presence of a  compliant substrate. By monitoring magnetotransport signals at low temperature, one can extract both spin--orbit length and coherence length, both crucial ingredients for the realization of more complex hybrid devices. As we will show, the presence of electrostatic gates allows for the external control of both lengths. Importantly, it can induce additional Rashba field, allowing to introduce and control vectorial spin--orbit interactions within the nanowire. 

Second, we report on the realization of hybrid Josephson junctions formed by an InAs quantum well between two Nb contacts\cite{Guiducci2018, Guiducci2019}. Here, we demonstrate supercurrent flow, whose amplitude can be tuned acting on two side gates, with Nb critical fields of the order of 3T. It has been shown that in these structures, at these magnetic fields, well-developed quantum Hall plateaus appear,\cite{Guiducci2018} allowing for the study of the coexistence of superconducting correlations and quantum Hall regime\cite{Mason2016,Stone2011,Ostaay2011}, two important ingredients for the emergence of topological quasiparticles in 2DEG-based hybrid structures.\cite{Alicea2016} As we will discuss, supercurrent flow through the device has been inspected by measuring multiple Andreev reflections (MAR). Device properties, such as carrier density, channel width and supercurrent amplitude, can be tuned by means of electrostatic gates. The presented data are in agreement with theoretical modeling, and are important step toward a complete characterization and understanding of hybrid semiconductor structures with strong spin--orbit coupling, in presence of external fields.

\section{Suspended nanowires}
\label{sec:nw}

\begin{figure}[tbp]
\begin{center}
\includegraphics[width=0.8\textwidth,keepaspectratio=true]{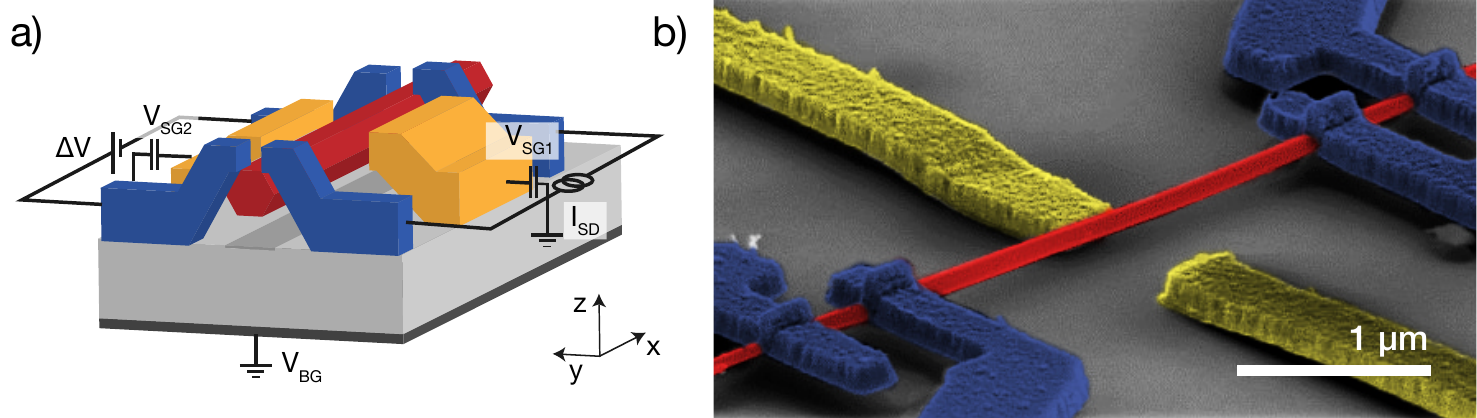}	
\caption{\label{fig1}(a) A sketch of the device with the four-wire measurement setup. A suspended InAs nanowire (in red) parallel to the $x$-axis is contacted with four ohmic contacts (in blue). The device lies in the $x-y$ plane. Two lateral gate electrodes (in yellow) allow to induce tunable electric fields inside the wire and modulate the SOI. (b) False-colored SEM image of a representative device.}
\end{center}
\end{figure}

Experimental evidences regarding the origin and vectorial nature of spin--orbit interactions (SOI) in semiconductor nanowires (NWs) and the vectorial dependence of the spin--orbit coupling are still limited, despite the growing literature and new phenomena, predicted and observed, related to its presence. Indeed, the confinement potentials would strongly affect spin--orbit coupling, and this unavoidably depends on the nanowire geometry and position with respect to the underlying substrate~\cite{degtyarev_features_2017,Bringer2011,nadj-perge_spectroscopy_2012,Roddaro2008}. To bypass spurious effects, freely suspended nanowires offer a unique platform and are not expected to display any intrinsic electrostatic asymmetry\cite{iorio_nanolett_2019}. In this way, one can restore the natural degeneracy of the system, allowing for the investigation  of intrinsic properties of spin--orbit interactions. The vectorial dependence of this quantity can be investigated with low-temperature magnetotransport, by tracking the weak anti-localization (WAL) peak~\cite{bergmann_weak_1984, chakravarty_weak_1986} and measuring the correction to magnetoconductance as a function of bias and magnetic field. Moreover, the presence of a pair of local side gates allows to tune the electron spin dynamics by exploiting the Rashba effect.

InAs nanowires under investigation were grown by Au-assisted chemical beam epitaxy following the procedure of Ref.~\cite{Gomes_2015}. The fabrication of the suspended structures followed a novel, and challenging, procedure with several steps based only on electron beam lithography techniques, which provide a deterministic procedure to realize suspended wires. A $\sim 200$~nm thick polymer PMMA sacrificial layer is first deposited over a Silicon wafer substrate, onto which the InAs nanowires are dropcasted after having suspended them in IPA. In order to provide the suspension, the PMMA is crosslinked at the two nanowire ends by exposing it to  high doses of electron beam ($5000$~$\mu$C/cm$^2$ at {$20$~keV). Next, the remaining, not crosslinked areas are dissolved. The crosslinked PMMA reduces its original thickness by 30-50\%, thus the wires result $\sim 100$~nm above the underlying substrate. Another PMMA layer is used in the final lithographic step in which metallic contacts and two side gates are patterned, defining the device structure as shown in Fig.~\ref{fig1}\cite{iorio_nanolett_2019}.

\begin{figure}[tbp]
\begin{center}
\includegraphics[width=0.8\textwidth,keepaspectratio=true]{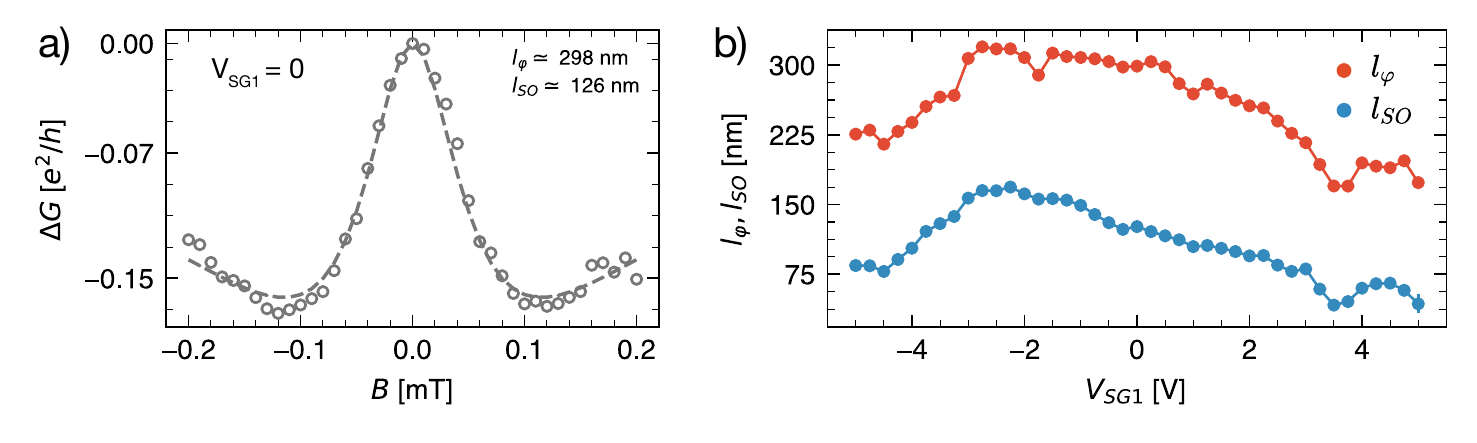}	
\caption{\label{fig2}(a) The conductance correction $\Delta G(B)=G(B)-G(0)$ is shown as a function of magnetic field applied along the $z$-axis at zero external gate voltages $V_{SG1}=V_{SG2}=0$. A clear peak at low field is due to the weak anti-localization effect. (b) The phase coherence length $l_\varphi$ and spin--relaxation length $l_{so}$ are shown as a function of the applied side gate potentials, asymmetrically swept as $V_{SG1}=-\alpha V_{SG2}$ with $\alpha = 0.4$.}
\end{center}
\end{figure}

The experimental results are obtained using four-probe transport measurements, in which a current $I_{SD}$ is applied between the source-drain electrodes and the corresponding voltage drop $\Delta V$ is measured by two other inner probes. This allows to obtain the nanowire resistance $R=\Delta V/I_{SD}$ avoiding the voltage drop over the contact resistances and setup filtering stages. Standard electrical characterization allows to extract nanowire parameters by monitoring the gate dependence of the differential conductance. We have obtained mobility $\mu\sim 10^3$~cm$^2$/(Vs), carrier density $n_{NW} \sim 3\cdot10^{18}$~cm$^{-3}$, and electron MFP of the order of $\ell\sim20-30$~nm. 

The device is fabricated with a pair of gates located on opposite sides of the nanowire. By placing an asymmetric potential $V_{SG1}=-\alpha V_{SG2}$ on the two gates, a tunable electric field can be induced inside the NW without changing its overall charge density. {The parameter $\alpha=0.4$ has been chosen to result in a constant NW transconductance and therefore takes into account the asymmetries of the two gates with respect to the wire axis.} This allows to attribute the tuning of Rashba spin--orbit strength only to the external electric field and not to a change in the carrier density or to a density--dependent scattering time, as previously argued in similar systems~\cite{miller_gate-controlled_2003, tikhonenko_transition_2009}. Additional Rashba spin--orbit contributions, induced by electrostatic field effect, can thus introduce vectorial spin--orbit components in the system.

In semiconductor nanowires, the presence of a limited amount of scattering centers gives rise to universal conductance fluctuations (UCFs) which can eventually mask localization features. However, different fluctuation patterns, which are determined by the specific impurity configuration, can be made accessible by changing the localization of the charge inside the wire through a back gate potential. Averaging techniques are needed in order to recover the localization effects, e.g. by averaging the conductance $G$ of a single wire on the different UCF patterns obtained by gating \cite{iorio_nanolett_2019,Jespersen2015,Jespersen2018}. In our case, this is achieved by using an AC modulation to the back gate with a low frequency voltage of $f = 10.320$ Hz and peak-to-peak amplitude $V_{bg}^{avg} = 6$ V.

A peak in the magnetoconductance at $B=0$ emerges after having averaged the signal, a clear hallmark of the weak anti-localization effect. This is displayed in Fig.~\ref{fig2}(a) which shows the conductance correction $\Delta G(B) = G(B)-G(0)$ measured at $T=50$ mK and at zero gate voltage $V_{SG1}=V_{SG2}=0$. By comparing the experimental data to the quasi-1D theory of WAL for disordered systems~\cite{kurdak_quantum_1992, altshuler_magnetoresistance_1981, iorio_nanolett_2019}, a phase coherence length $l_\varphi \sim 300$ nm and spin--relaxation length $l_{so}\sim 125$ nm can be estimated. Since no external electrostatic asymmetry is present in the suspended structure, the observed WAL can be attributed to the internal electric field due to the Fermi level pinning at the NW surface~\cite{iorio_nanolett_2019}.

By monitoring the WAL for different values of the side gate potentials we are able to modulate the spin--relaxation length as a function of the applied external fields. This is depicted in Fig.~\ref{fig2}(b) which shows the behavior of $l_\varphi$ and $l_{so}$ as a function of the asymmetrically swept side gate potentials. A reduction of the spin--relaxation length by almost a factor 3 is obtained for $V_{SG1}\sim 4$ V with $l_{so} \sim 50$ nm, as a consequence of the enhancement of the SOI inside the NW acting through the Rashba effect. The intrinsic NW confining potential, preserved in the suspended structure, can be responsible for the small asymmetry in the observed modulation which shows a peak at $-2$~V, that can be attributed to the competing effect of a slightly tilted external electric field and the internal NW potential. The same modulation of $l_{so}$ is also visibile for the coherence length $l_\varphi$, as frequently reported in similar works\cite{scherubl_electrical_2016, wang_electrical_2017}. Although an enhanced electron-electron inelastic scattering rate could justify the observed trend, the origin of the strong correlation between these two quantities is still under debate.

In conclusion, we have investigated the electrical control of SOI in suspended NWs. A strong reduction of the spin--relaxation length by almost a factor 3 is achieved at a constant NW conductance. The observed modulation confirms that the Rashba effect, both triggered by internal or external electric fields, is the main source of SOI in InAs NWs. The achieved electrical control can be exploited for future spintronic applications and for the manipulation of Majorana bound states.

\section{Hybrid Josephson junctions}
\label{sec:scqh}

\begin{figure}[tbp]
\begin{center}
\includegraphics[width=0.8\textwidth,keepaspectratio=true]{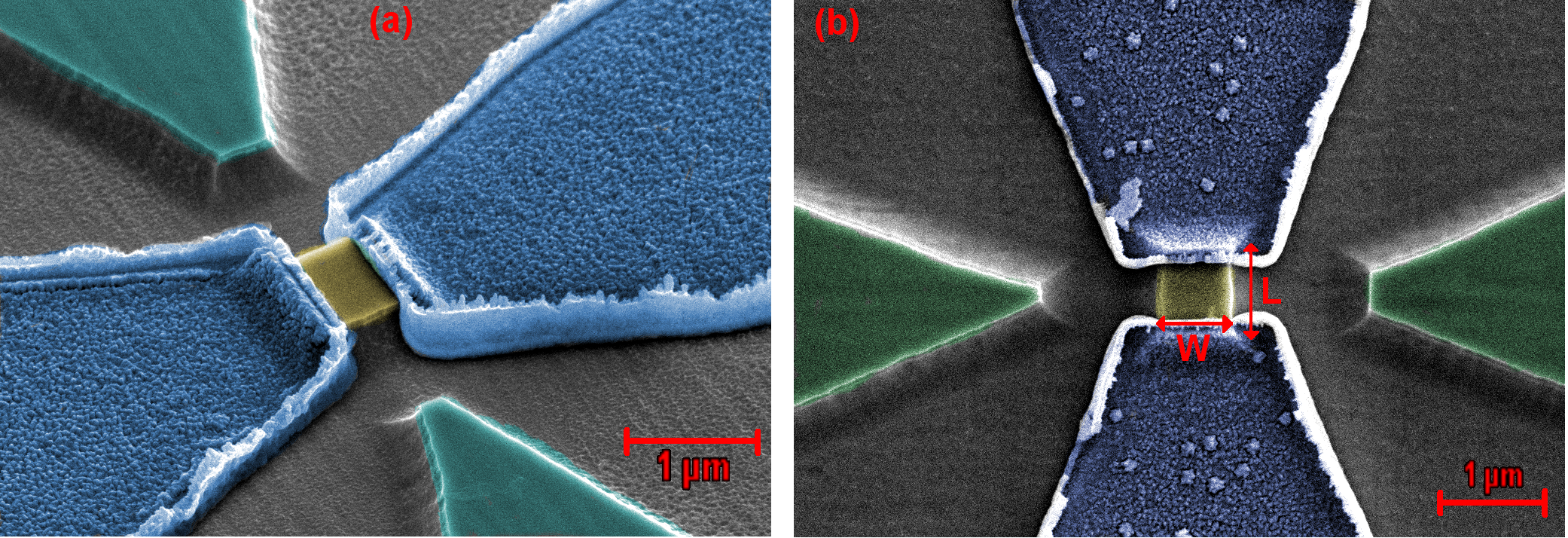}
\caption{\label{picture:Devices1}(a) False color SEM image of a tilted device. The mesa is yellow, side gates are green, niobium is blue. (b) Top view of a device. Length ($L$) and width ($W$) of the mesa are shown with red arrows.}
\end{center}
\end{figure}

Hybrid Josephson junctions comprising a two dimensional electron gas in the normal region have been recently investigated\cite{Amet2016,Lee2017,Shalom2016,Wu2018,Guiducci2018}, especially looking for signatures of the coexistence of superconductivity and quantum Hall regime\cite{Alicea2016, Mason2016,Stone2011,Ostaay2011, Ferraro_2014}. Here, we consider Josephson junction devices formed with a high-quality InAs quantum-well contacted by two Nb leads. A high quality InAs quantum well was grown by molecular beam epitaxy on a GaAs substrate\cite{biasiol1, biasiol2, Amado2014, Guiducci2018, Guiducci2019}, as detailed in Refs.\cite{Amado2014, Guiducci2018}. Carrier concentration of the 2DEG is $n_{2d} = 6.2\cdot 10^{11}$~cm$^{-2}$, electron mobility $\mu=1.6\cdot 10^5$~cm$^2$/(Vs), and electron MFP $\ell=2.2$~$\mu$m. Two superconducting Nb leads are placed by multiple electron beam lithography steps (see \cite{Amado2014, Guiducci2018} for details), as shown in Fig.~\ref{picture:Devices1}, forming  the hybrid Josephson junction.  Additional side gates were patterned, in another lithographic step, close to the mesa region, allowing for the tuning of both carrier density and channel width of the normal region. Nb leads laterally contact the InAs quantum well, as shown in Fig.~\ref{picture:Devices1}, and several gold Ohmic pads are present allowing for four-probe transport measurements on the Nb. Experiments were performed at $T\sim 300$~mK with appropriate filtering of all transport lines\cite{Guiducci2018}. Temperature evolution of source--drain current in a two--probe ac configuration allowed the extraction of the Nb critical temperature with a value of $T_c=8.7$~K, with an associated gap $\Delta_0 \sim 1.33$~meV. Supercurrent flow, or the Josephson regime, can be probed in a current--bias configuration by monitoring the source--drain voltage response, and is characterized by a flat and zero voltage central region. From these measurements, a critical current $I_c\sim 170 $~nA has been reported, which can be tuned down to zero by acting on the external side gates, forming a so-called Josephson field effect transistor\cite{Guiducci2018, Akazaki1996,Clark1980,Bezuglyi2017}.

Important, and complementary, information can be obtained also in the dissipative regime, where a finite source-drain voltage $V_{SD}$ develops. In particular, sub--gap structures can be analyzed looking at the differential resistance at voltage values $V_{SD}\leq 2\Delta_0/e$. During these current--bias measurements, the ac component $I_{ac}$ has been fixed, while the dc component $I_{dc}$ is swept from zero to $I_{dc}=-3.3\mu$A, and the ac and dc components ($V_{ac}$ and $V_{dc}$, respectively) of $V_{SD}$ are measured with two different instruments. From the former, one directly gets the value of the differential resistance, while the latter corresponds to $V_{SD}$. Figure~\ref{picture:MAR1}(a) shows differential resistance as a function of source-drain voltage $V_{SD}$ for a representative device. At $V_{SD}=0$~V, the differential resistance drops to zero; this is the superconductive state. The peak next to the superconductive state is due to the sudden jump between the superconductive state and the normal state. Conversely, at $V_{SD}\approx -3$~mV, the differential resistance is equal to the normal resistance ($R_N$). Between these two extrema there are several oscillations, which represent the sub--gap structures expected for a working Josephson junction. These sub--gap structures can be analyzed in detail by looking for multiple Andreev reflections, i.e. local minima of resistance. According to standard theory\cite{obtk,Tinkham,Heida1998}, these regions follow
\begin{equation}\label{eq:MARs2}
    V_n=\frac{2\Delta_{0}}{en}, \quad \quad n = 1, 2, 3... .
\end{equation}
The straight lines in Fig.~\ref{picture:MAR1}(b) show the series from $n=1$ to $n=7$, with an excellent agreement with the theoretical expectation (see in particular $n=1$ to $n=5$). From this measurement, we extract the value of $\Delta_0$, obtaining $\Delta_0=1.37\pm0.03$~meV, in excellent agreement with the value obtained from the critical temperature $T_c$. Figure~\ref{picture:MAR1}(c) shows similar data for a different range of source-drain voltages. It reports another series of MARs with $n=1$, $2$, and $3$. Now, the extracted gap value is $\Delta^*=(125 \pm 10)$~$\mu$eV. Such energy value is called minigap, i.e.~a small energetic gap, induced by the superconductor in the density of states of the 2DEG thanks to proximity~\cite{pannetier2000, courtois1999, belzig1999, virtanen2017, bours18, vischi2017}. The value of the proximity induced gap found here is consistent with the ones reported in similar hybrid structures\cite{Deon2011, Irie2014}.

\begin{figure}[tbp]
\begin{center}
\includegraphics[width=0.8\textwidth,keepaspectratio=true]{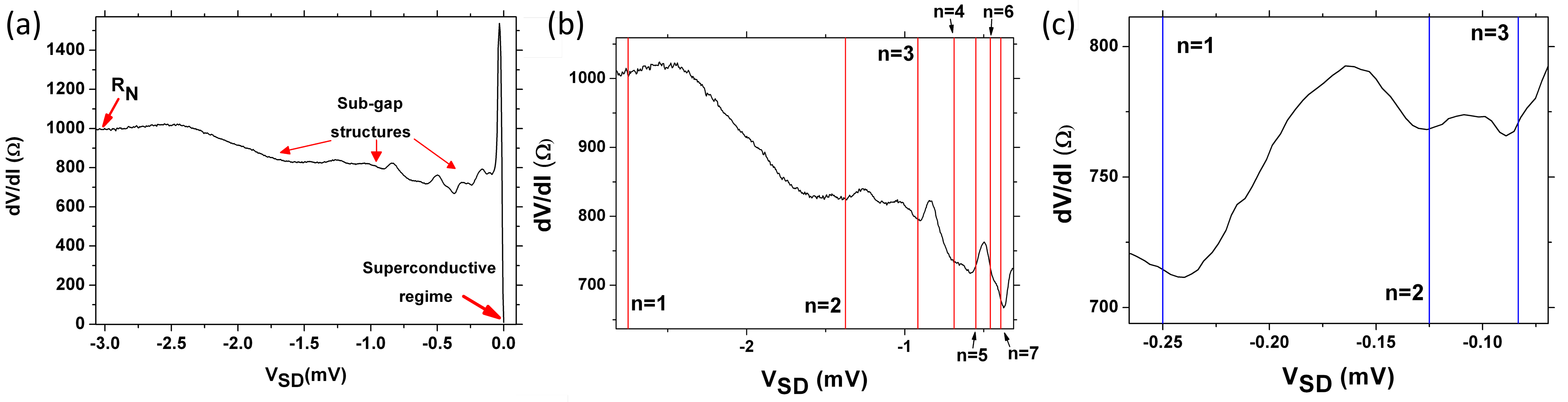}
\caption{\label{picture:MAR1}Differential resistance $\frac{dV}{dI}$ vs.~source-drain voltage $V_{SD}$. (a) The superconductive regime, the normal resistance ($R_N$), and the sub--gap regions are indicated by arrows. (b) Series $V_n=\frac{2\Delta_{Nb}}{en}$ from $n=1$ to $n=7$ (red lines). (c) Series $V_n=\frac{2\Delta^*}{en}$ from $n=1$ to $n=3$ (blue lines).}
\end{center}
\end{figure}

Figure \ref{mar2} shows the differential resistance of another device taken at various external magnetic fields $B$. Figure \ref{mar2}(a) shows the differential resistance versus $V_{SD}$ for $B=0.2$~mT (blue) and $B=4$~mT (black). The series from Eq.~\ref{eq:MARs2} for the Nb gap is plotted from $n=1$ to $n=4$ (black labels), and the series for the minigap $\Delta^*$ from $n=1$ to $n=3$ (red labels). For this device, $\Delta^*=(220 \pm 20)$~$\mu$eV. This time the niobium series is less sharp, but minima for $n=1$, $n=3$, and $n=4$ are visible. At $B=0.2$~mT, the measurement shows three additional sharp minima below $|V_{SD}|=0.5$~mV, correctly intersected by the $\Delta^*$ series. These three minima are due to $\Delta^*$ and are suppressed by the magnetic field; in fact, at $B=4$~mT they are not visible anymore. On the other hand, the niobium minima almost do not change with magnetic field. Figure~\ref{mar2}(b) shows the same measurement as before, but for a larger number of magnetic field values ($41$ instead of $2$). The plot shows  differential resistance ($z$-axis: colors) as a function of $V_{SD}$ ($x$-axis) and magnetic field $B$ ($y$-axis). The series $V_{SD}=\frac{2\Delta^*}{en}$  is indicated by red labels. Minima are centered at $B=0.2$~mT and at $V_{SD}=\frac{2\Delta^*}{en}$ for $n=1, 2,\text{ and } 3$; but they rapidly vanish when departing from $B=0.2$~mT (in both directions). They disappear approximately at $2$~mT; this means that the minigap is suppressed above these magnetic fields. On the contrary, in order to suppress the niobium gap, much higher magnetic fields are necessary (of the order of 3~T)\cite{Guiducci2018, Guiducci2019}, and in fact the niobium series is stable and does not change in the range of fields considered here. These measurements demonstrate that the devices work properly as Josephson junctions, with supercurrent flow, and provide evidence of sub--gap structures, which well agree with multiple Andreev reflections expected by a simple theoretical model. From these measurements, the value of the minigap can be extracted, showing that a proximity effect within the 2DEG has been induced by the Nb leads.

\begin{figure}[tbp]
\begin{center}
\includegraphics[width=0.8\textwidth,keepaspectratio=true]{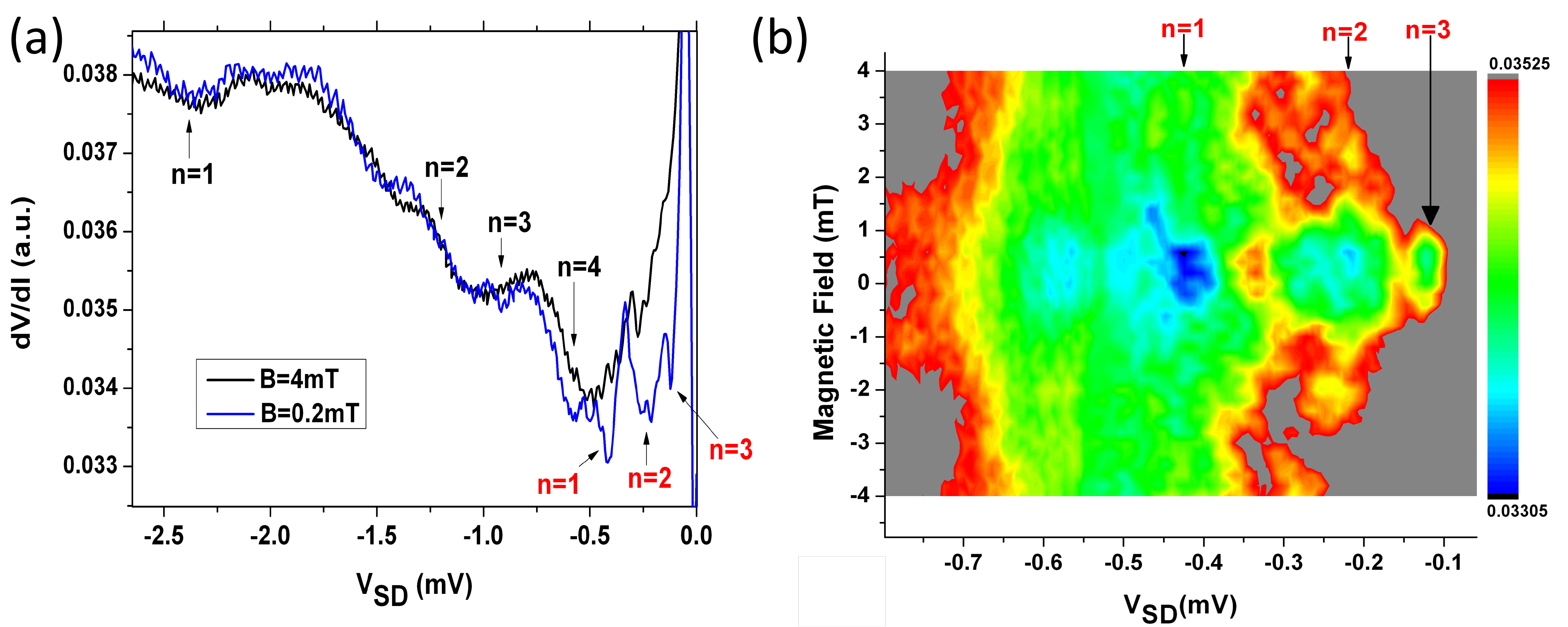}
\caption{\label{mar2}(a) Differential resistance (arbitrary units) vs.~$V_{SD}$ for $B=0.2$~mT (blue) and $B=4$~mT (black). Niobium series (with black labels) and $\Delta^*$ series (with red labels) are also plotted. (b) 3D plot showing the differential resistance (colors) vs.~both magnetic field (from $-4$~mT to $4$~mT) and $V_{SD}$ (from $-0.05$~mV to $-0.8$~mV). $\Delta^*$ series is shown with red labels on top of the plot. The two curves shown in (a) are horizontal lines in (b) at $B=0.2$~mT and $B=4$~mT. Performed at $V_{gate}=0$~V and $T=362$~mK.}
\end{center}
\end{figure}

\section{Summary}
\label{sec:conclusions}

Semiconductors with strong spin--orbit coupling are of great interest for their potential in a wide range of applications. InAs--based devices are currently subject of intense research, with particular emphasis on hybrid superconductor/semiconductor structures. Here we have reported on two different setups whose physics has been investigated by low-temperature transport measurements. In the first case, a suspended InAs nanowire has been considered, obtaining important information on the intrinsic properties related to spin--orbit interactions. Moreover, we have shown that modulation of spin--orbit length can be obtained by properly acting on two external side gates. In the second case, we have focused on hybrid Josephson junctions formed by an InAs quantum-well placed between two Nb contacts. We have observed multiple Andreev reflections, from which we have extracted important information on gap values and proximity induced minigap. Also in these devices, the presence of external side gates represents a useful knob to tune observable quantities such as supercurrent amplitude.   

\section*{Acknowledgments}
M.C., V. Z., and L.S. acknowledge support from the Quant-EraNet project Supertop. S.G. has been supported by Fondazione Silvio Tronchetti Provera. A.I., E.S., and F.G. acknowledge support from the Horizon research and innovation programme under grant agreement No. 800923 (SUPERTED). M. R. has received funding from the European Union's Horizon 2020 research and innovation programme under the Marie Sklodowska-Curie grant agreement
EuSuper No 796603.
 
\bibliography{bibliography} 
\bibliographystyle{spiebib} 

\end{document}